\documentclass[prd,aps,showpacs]{revtex4}

\usepackage{epsfig}
\usepackage{graphicx,color}

\newcommand{\beq}{\begin{equation}}
\newcommand{\eeq}{\end{equation}}
\newcommand{\be}{\begin{eqnarray}}
\newcommand{\ee}{\end{eqnarray}}

\begin{document}

\title{
Direct search for heavy gauge bosons at the LHC
in the nonuniversal SU(2) model 
}

\author{ Yeong Gyun Kim }

\affiliation{
Department of Science Education,
Gwangju National University of Education, Gwangju 500-703, Korea
}

\author{ Kang Young Lee }
\email{kylee.phys@gnu.ac.kr}

\affiliation{
Department of Physics Education \&
Research Institute of Natural Science,
\\
Gyeongsang National University, Jinju 660-701, Korea
}

\date{\today}

\begin{abstract}

We study the phenomenology of heavy gauge bosons at the LHC
in a nonuniversal gauge interaction model
with the separate electroweak SU(2) gauge group for the third generation.
Considered are the Drell-Yan processes into the final states 
of dilepton, dijet, $\tau^- \tau^+$, and $t \bar{t}$
for $Z'$ boson and those of lepton-neutrino for $W'$ boson.
We find that the present LHC data provides lower bounds 
on the masses of the heavy gauge bosons, $m_{Z'},~m_{W'} > 2$ TeV, 
more stringent than indirect bounds,
when $\sin^2 \phi > 0.15$ 
for the mixing angle of two SU(2) gauge groups.
We also note that the study of heavy resonances into the third generation
fermions may provide some valuable information in the future. 

\end{abstract}


\maketitle

\section{Introduction}

The CERN Large Hadron Collider (LHC) has 
successfully completed the first stage of running
with the discovery of a standard model (SM)-like Higgs boson.
Still there is no evidence
for the new physics beyond the SM at the LHC so far.
We have to wait till the next stage of the LHC
in order to find new results.
However, we can read out valuable informations
on the new physics scales from the present data.
One of the most promising signal of new physics beyond the SM
is the heavy resonances decaying into a pair of the SM particles. 
The CMS \cite{cmsll,cmsjj,cmstt,cmstautau,cmslnu} and 
ATLAS \cite{atlasll,atlasjj,atlastt,atlastautau,atlaslnu} collaborations 
has reported the search results for the extra gauge bosons, $W'$ and $Z'$,
with the data collected at the LHC in 2011 and in 2012.
Recent search results at the LHC shows the absence of heavy mass resonances
and present strong bounds on $m_{W'}$ and $m_{Z'}$ more than 2 TeV.

Many new physics models in the context of the gauge unification
contain extended gauge symmetry and 
predict the existence of extra neutral and/or charged gauge bosons 
with heavy masses.
It is very important to search for the extra gauge bosons directly
and to study their phenomenology at the LHC in detail
\cite{extra_gauge}.
In this paper, we consider an extended model for the electroweak gauge group.
The SM assumes the universality of the electroweak symmetry
on all the fermion generations.
If a separate SU(2) symmetry acts on the third generation, 
the nonuniversal nature of the electroweak symmetry 
as well as the additional gauge bosons
presents various interesting phenomenology
\cite{topflavor,lee1}.
Such a gauge group might be vestiges of the family symmetry
or a symmetry at an intermediate stage in the path of 
symmetry breaking of noncommuting extended technicolor models
\cite{ETC}. 
We assign the left-handed quarks and leptons 
for the first and second generations $(2,1,1/3)$, $(2,1,-1)$
and those for the third generation $(1,2,1/3)$, $(1,2,-1)$
under SU(2)$_l \times$SU(2)$_h \times$ U(1)$_Y$.
In the same manner 
the right-handed quarks and leptons transform as (1,1,2$Q$)
with the electric charge $Q=T_3^l + T_3^h + Y/2$.
The separate SU(2) is mixed with the ordinary SU(2) in general
and should be broken to the SM gauge group at a high energy scale $u$.
It can be achieved by introducing an bidoublet scalar field $\Sigma$ (2,2,0)
with the vacuum expectation values (VEV) 
$ \langle \Sigma \rangle = {\rm Diag} (u,u)$. 
The electroweak symmetry breaking is performed 
by an additional scalar field at the scale $v$.
The detailed discussion of the Higgs sector in this model
can be found in Ref. \cite{chiang}.
The phenomenology of this model has been studied 
using the low-energy data
\cite{topflavor,lee1,lee2,malkawi2}.
The nonuniversality of this model derives the exotic flavor-violating 
terms in both neutral and charged current interactions.
They give rise to the lepton flavor violations and the violation 
of unitarity of the CKM matrix,
which lead to strong constraints on the model parameters
\cite{ckm,lfv}.

Being introduced an additional SU(2) gauge symmetry,
extra charged and neutral gauge bosons $W'$ and $Z'$ with heavy masses
exist in this model.
In this work, we study the phenomenology of $W'$ and $Z'$
with the data collected at the first run of the LHC.
The direct bound on the $W'$ boson mass has been obtained
from the early data of the LHC in the ref. \cite{lee3}.
Here we update the $W'$ data and 
perform the new analyses on the various channels of $Z'$ boson.
The $W'$ and $Z'$ decays will be considered 
through the Drell-Yan mechanism which is the $s-$channel process 
mediated by $W'$ or $Z'$ into the fermion pairs.
We obtain direct lower bounds on the $W'$ and $Z'$ masses and 
constraint on the mixing angle between two SU(2) groups
from the lack of the signal of heavy gauge boson at the LHC.
In the next section, we will briefly review the model and discuss
the phenomenology of the heavy gauge bosons.
The analysis of the LHC results in this model is presented
in section III. Finally we conclude.

\section{Phenomenology of heavy gauge bosons in the nonuniversal SU(2) model}

After the gauge symmetry in this model is broken to the SM gauge group 
and sequently to the U(1)$_{\rm EM}$, 
we parameterize the gauge coupling constants as
\be
g_l = \frac{e}{\sin \theta \cos \phi},~~~~
g_h = \frac{e}{\sin \theta \sin \phi},~~~~
g^{\prime} &=& \frac{e}{\cos \theta},
\ee
in terms of the electromagnetic coupling $e$, 
the weak mixing angle $\theta $ and the new mixing angle $\phi$ 
between $SU(2)_l$ and $SU(2)_h$.
In this analysis, we assume the perturbativity of
all of the gauge couplings, $g_{(l,h)}^2/4 \pi < 1$,
which is corresponding to $ 0.03 < \sin^2 \phi < 0.96 $.

For simplicity of the analysis, 
we introduce a small parameter $ \lambda \equiv v^2/u^2 $.
We define the physical state of gauge bosons $W'$ by
\be
\left( \begin{array}{c}
W_\mu^\pm\\
{W'}_\mu^\pm
 \end{array} \right)
&=&
\left( \begin{array}{cc}
1 & \lambda \sin^3 \phi \cos \phi\\
-\lambda \sin^3 \phi \cos \phi & 1
 \end{array} \right)
\left( \begin{array}{cc}
\cos \phi & \sin \phi \\
-\sin \phi &  \cos \phi
 \end{array} \right)
\left( \begin{array}{c}
{W_l}_\mu^\pm\\
{W_h}_\mu^\pm
 \end{array} \right),
\ee
and those of $Z'$ by
\be
\left( \begin{array}{c}
Z_\mu\\
{Z'}_\mu\\
A_\mu
 \end{array} \right)
&=&
\left( \begin{array}{ccc}
1 & \lambda \frac{\sin^3 \phi \cos \phi}{\sin \theta} & 0\\
-\lambda \frac{\sin^3 \phi \cos \phi}{\sin \theta} & 1 & 0 \\
0 & 0 & 1
 \end{array} \right)
\left( \begin{array}{ccc}
\cos \theta \cos \phi & \cos \theta \sin \phi & -\sin \theta \\
-\sin \phi &  \cos \phi & 0 \\
\sin \theta \cos \phi & \sin \theta \sin \phi & \cos \theta \\
 \end{array} \right)
\left( \begin{array}{c}
{W_l^3}_\mu\\
{W_h^3}_\mu\\
B_\mu
 \end{array} \right)~,
\ee
in the leading order of $\lambda$.
Their masses are given by 
\be
m_{W'}^2 = m_{Z'}^2
= \frac{m_0^2}{\lambda \sin^2 \phi \cos^2 \phi} 
  \left( 1 + {\cal O(\lambda)} \right) ,
\ee
where $m_0 = ev/(2 \sin \theta)$ is 
the ordinary $W$ boson mass at tree level.
We note that the $W'$ and $Z'$ masses are degenerate in this model.

Two independent parameters $(\lambda, \sin^2 \phi)$ are introduced 
to describe the new physics effects in this model. 
We keep the linear order of the small parameter $\lambda$ in this paper.
Presenting the results of phenomenological analyses, 
we will use the observable quantity $m_{Z'} (=m_{W'})$ instead of $\lambda$
as a model parameter.

We derive the neutral current interaction for $Z'$ boson such that
\be
{\cal L}_{NC} = {G'}_L \bar{f}_L \gamma_\mu {Z'}^\mu f_L 
              + {G'}_R \bar{f}_R \gamma_\mu {Z'}^\mu f_R 
              + {X'}_L \bar{f}_L \gamma_\mu {Z'}^\mu f_L, 
\ee
where
\be
{G'}_L &=& \frac{e}{\sin \theta} \tan \phi~(T_{3l}+T_{3h}) + {\cal O}(\lambda),
\nonumber \\
{G'}_R &=& {\cal O}(\lambda),
\nonumber \\
{X'}_L &=& -\frac{e}{\sin \theta} \frac{1}{\sin \phi \cos \phi} ~T_{3h} 
           + {\cal O}(\lambda).
\ee
Note that ${G'}_L$ and ${G'}_R$ are universal couplings
and ${X'}_L$ are the couplings only for the third generations.
The charged current interactions for $W'$ boson are also given by
\be
{\cal L}_{CC} = V_{UD} \bar{U}_L \gamma_\mu {H'}_L {W'}^\mu D_L 
              + V_{UD} \bar{U}_L \gamma_\mu {Y'}_L {W'}^\mu D_L 
+ {\rm H.~c.},
\ee
for quarks where $U_L=(u_L, c_L, t_L)^T$, $D_L =(d_L, s_L, b_L)^T$
and 
\be
{H'}_L &=& -\frac{g}{\sqrt{2}} \tan \phi,
\nonumber \\
{Y'}_L &=& -\frac{g}{\sqrt{2}} \frac{1}{\sin2 \phi \cos \phi} {\hat Y}_3,
\ee
where ${\hat Y}_3$ is a $3\times 3$ matrix 
with elements $\delta_{i3} \delta_{j3}$.
We note that the CKM matrix is also shifted by ${\cal O}(\lambda)$ terms,
which is severely constrained by the precise test 
of the CKM matrix unitarity \cite{ckm}.
We let the matrix elements $V_{UD}$ to be the SM values in this work
to keep the decay rates in the leading order. 

\begin{figure}[t!]
\centering
\includegraphics[height=8cm]{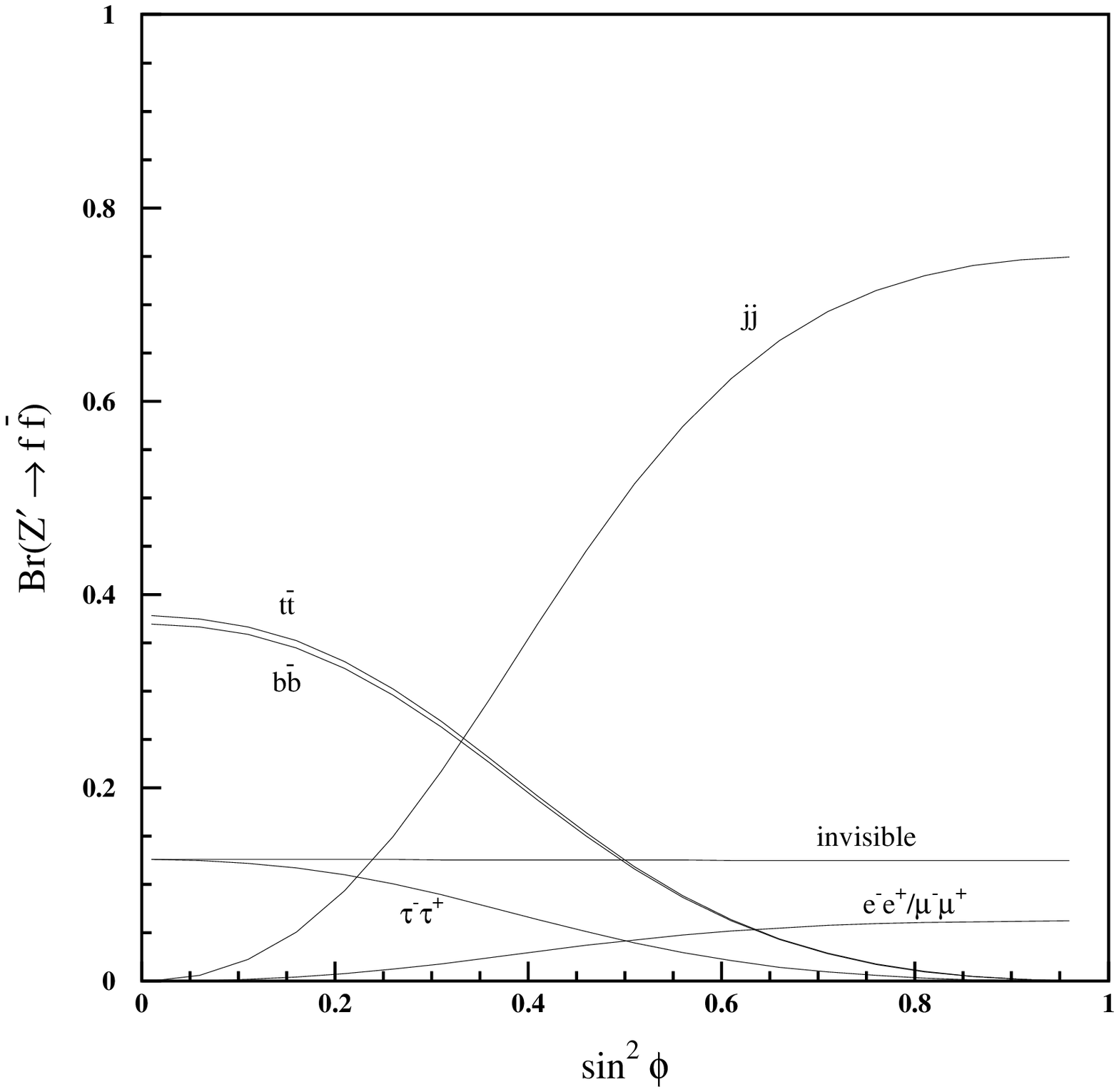}
\includegraphics[height=8cm]{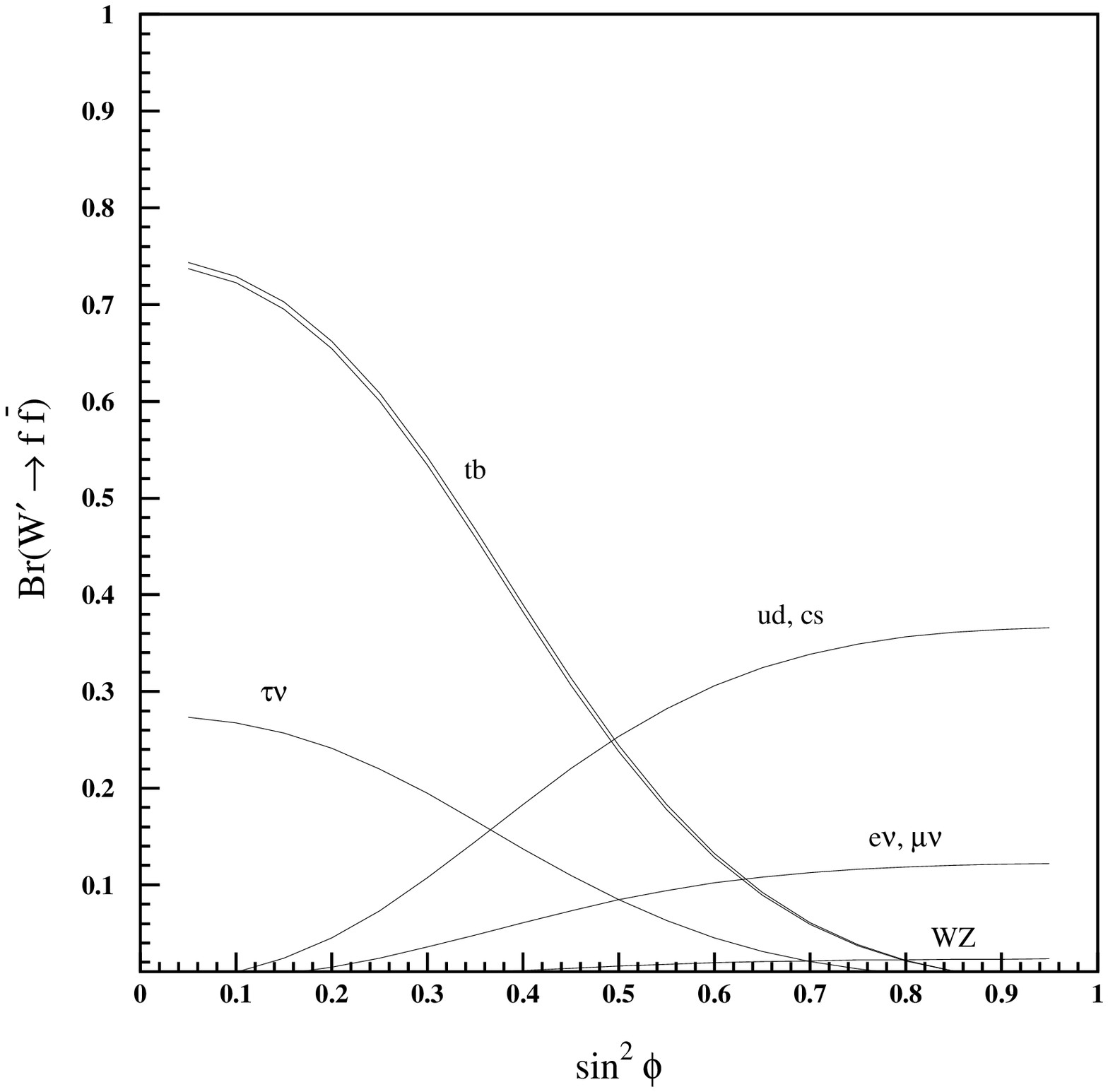}
\caption{
Branching ratios of the $Z'$ and $W'$ boson with respect to $\sin^2 \phi$.
}
\end{figure}

We obtain the decay rates of $Z'$ and $W'$ bosons  
from the replacements of the ordinary couplings and masses
by those of heavy gauge bosons 
in the SM decay rates of $Z$ and $W$ boson. 
We have
\be
\Gamma (Z' \to f \bar{f}) &=& \Gamma_0^Z \frac{m_{Z'}}{m_Z} \cdot \tan^2 \phi,
\nonumber \\
\Gamma (W' \to f \bar{f}') &=& \Gamma_0^W \frac{m_{W'}}{m_W} \cdot \tan^2 \phi,
\ee
for the first and second generations and
\be
\Gamma (Z' \to f \bar{f}) 
        &=& \Gamma_0^Z \frac{m_{Z'}}{m_Z} 
     \cdot \tan^2 \phi \left(1-\frac{1}{\sin^2 \phi} \right)^2,
\nonumber \\
\Gamma (W' \to f \bar{f}') 
        &=& \Gamma_0^W \frac{m_{W'}}{m_W} 
     \cdot \tan^2 \phi \left(1-\frac{1}{\sin^2 \phi} \right)^2,
\ee
for the third generation fermions
where 
$\Gamma_0^{Z}$ and $\Gamma_0^{W}$ are corresponding $Z$ and $W$
decay rates into the same final states in the SM.
You can see that even the top quark mass effects shows 
just a very small splitting.
Thus final state masses are ignored in this analysis.

Note that the triple gauge boson couplings involving $W'$ and $Z'$ 
arise in this model and
$W'$ and $Z'$ can decay into a pair of gauge bosons
through the triple gauge couplings.
Generically their decay rates are suppressed by small mixing angles.
Although the decays of the longitudinal modes of $W'$ and $Z'$ 
might becomes sizable by compensation of the factor 
of order $m_{W'}^4/m_W^4$,
the branching ratio of $W_L^{\prime} \to W_L Z_L$ is 
smaller than that of $W' \to e \nu$ 
by a numerical factor $ \cos^4 \theta_W/4$ 
and just less than 2$\%$. 
Moreover no direct search results are given for those channel yet
and we do not involve the processes with the gauge boson final states
in our analysis.

The branching ratios of $Z'$ and $W'$ boson are depicted in Fig. 1.
They do not depend on the heavy gauge boson masses but only on $\phi$ 
when the final state masses are ignored. 
Only the $Z' \to t \bar{t}$ and $W' \to tb$ decays show
a small splitting depending on $Z'$ and $W'$ masses 
due to the top quark mass effects. 
The decays into the third generation fermions dominate
in the small $\phi$ region,
while those rates are small in the large $\sin^2 \phi$ region.
It is because $W'$ and $Z'$ bosons are almost $W_h$ and $W_h^3$ bosons 
in the small $\phi$ region
and coupled to the third generations dominantly.
When $\sin^2 \phi \to 1$, 
$W'$ and $Z'$ mostly consist of $W_l$ and $W_l^3$ respectively
to decay into the first and second generations.

\begin{figure}[t!]
\centering
\includegraphics[width=16cm]{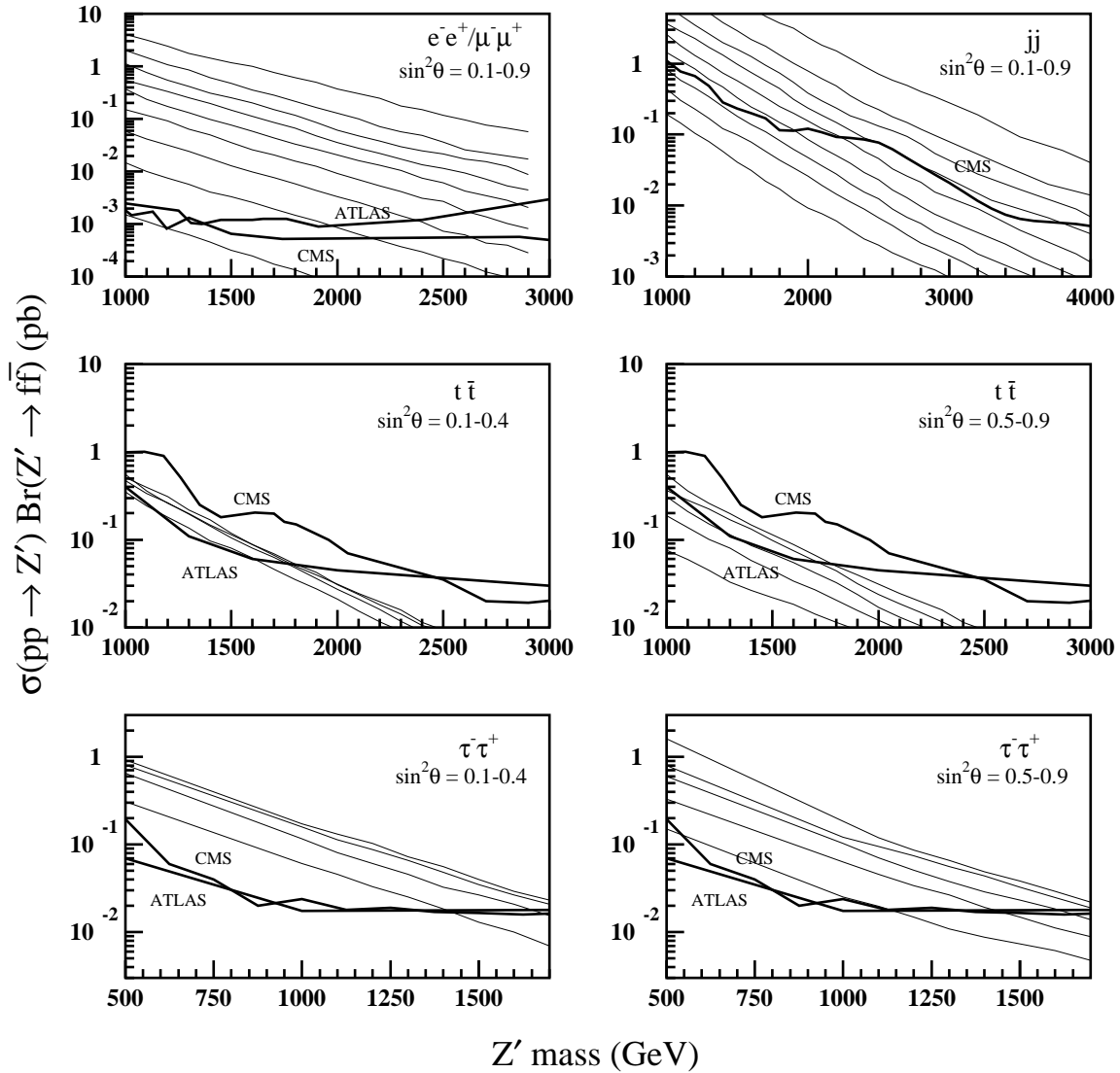}
\caption{
The thick lines denote the upper limits 
on the production cross section times branching ratios 
into various final states for {Z'} boson.
Above regions are excluded by the absence of the heavy resonance signatures.
The thin lines denote the theory predictions in this model. 
The top-left panel is for dilepton final states, the top-right for dijets,
the middle panels for $t \bar{t}$, and the bottom panels for $\tau^- \tau^+$.
All plots are for the combination of the 7 and 8 TeV data sets. 
}
\end{figure}

\section{Direct search at the LHC}

\begin{figure}[t!]
\centering
\includegraphics[height=8cm]{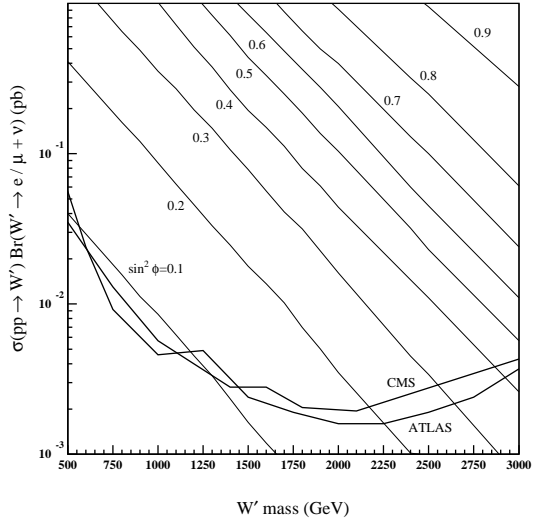}
\caption{
Cross sections times branching ratios into lepton-neutrinos for $W'$ boson
together with the updated experimental upper limits 
from the combined data at 7 and 8 TeV. 
}
\end{figure}

In order to find new resonances at the LHC, 
the most promising channels are 
dilepton and dijet final states for $Z'$ 
and lepton-neutrino channels for $W'$.
Processes involving the third generation fermions final states
might be important for search for the new physics models
violating universality.
We consider the dilepton, the dijet, $\tau^- \tau^+$, $t \bar{t}$
final states for $Z'$ searches and
$e \nu / \mu \nu$ final states for $W'$ searches.
The CMS and the ATLAS groups have measured the upper limit 
on the production cross section
times branching ratios for each channel.
The thick lines of Fig. 2 and 3 denote the experimental limits
from the data collected by the CMS and ATLAS collaborations 
in 2012 and in 2013 in part.
We calculate the production cross sections 
for $Z'$ and $W'$ gauge bosons
by using PYTHIA 6.4 \cite{PYTHIA}.
The theoretical predictions of the cross sections times branching ratios 
are shown as thin lines in Fig. 2 and 3 
together with the experimental limits from the LHC data. 
Each thin lines of the top panel in Fig. 2 
denote the predictions with $\sin^2 \phi=0.1$ to $0.9$
from the bottom to top. 
For the middle and bottom panels in Fig. 2,
the thin lines denote the predictions of $\sin^2 \phi=0.1$ to $0.4$
from the bottom to top for the left panels and 
of $\sin^2 \phi=0.5$ to $0.9$ from the top to bottom for the right panels.
The middle and bottom panels are for the processes
including the third generation fermions
and the production and decay processes show reverse behaviors
with respect to $\sin^2 \phi$.
Thus their product has the maximum at $\sin^2 \phi=0.5$ around.
Figure 3 depicts the updated analysis of Ref. \cite{lee3} 
on direct $W'$ search with the recent CMS and ATLAS data sets.
Since the regions above the thick lines are excluded at 95 \% C.L.,
we determine the direct lower bounds on the $W'$ and $Z'$ masses 
with respect to $\sin^2 \phi$ for each channel.

We present the direct search limits on $m_{Z'}$ and $m_{W'}$
together with indirect limits of the previous analysis. 
Indirect studies of this model consists on 
search for new physics signals in the neutral current interactions 
including the LEP and SLC data and the atomic parity violation (APV)
\cite{topflavor,lee1,malkawi2}
and search for non-SM signatures due to nonuniversality of
the SU(2) gauge interactions \cite{ckm,lfv}.
The nonuniversality provides the stronger constraints
on the model than the neutral current interactions data.
We see the direct search limits of lepton final states 
give the most stringent bounds except for very small
$\sin \phi$ region in Fig. 4.
As discussed above, the decays into the third generation fermions 
are dominant in the small $\phi$ region,
We note that the limit from $Z' \to \tau^- \tau^+$ channel
is relatively stronger in Fig. 4
and expect that this process will play an important role
with more data in the future run of the LHC.  

\begin{figure}[t!]
\centering
\includegraphics[height=8cm]{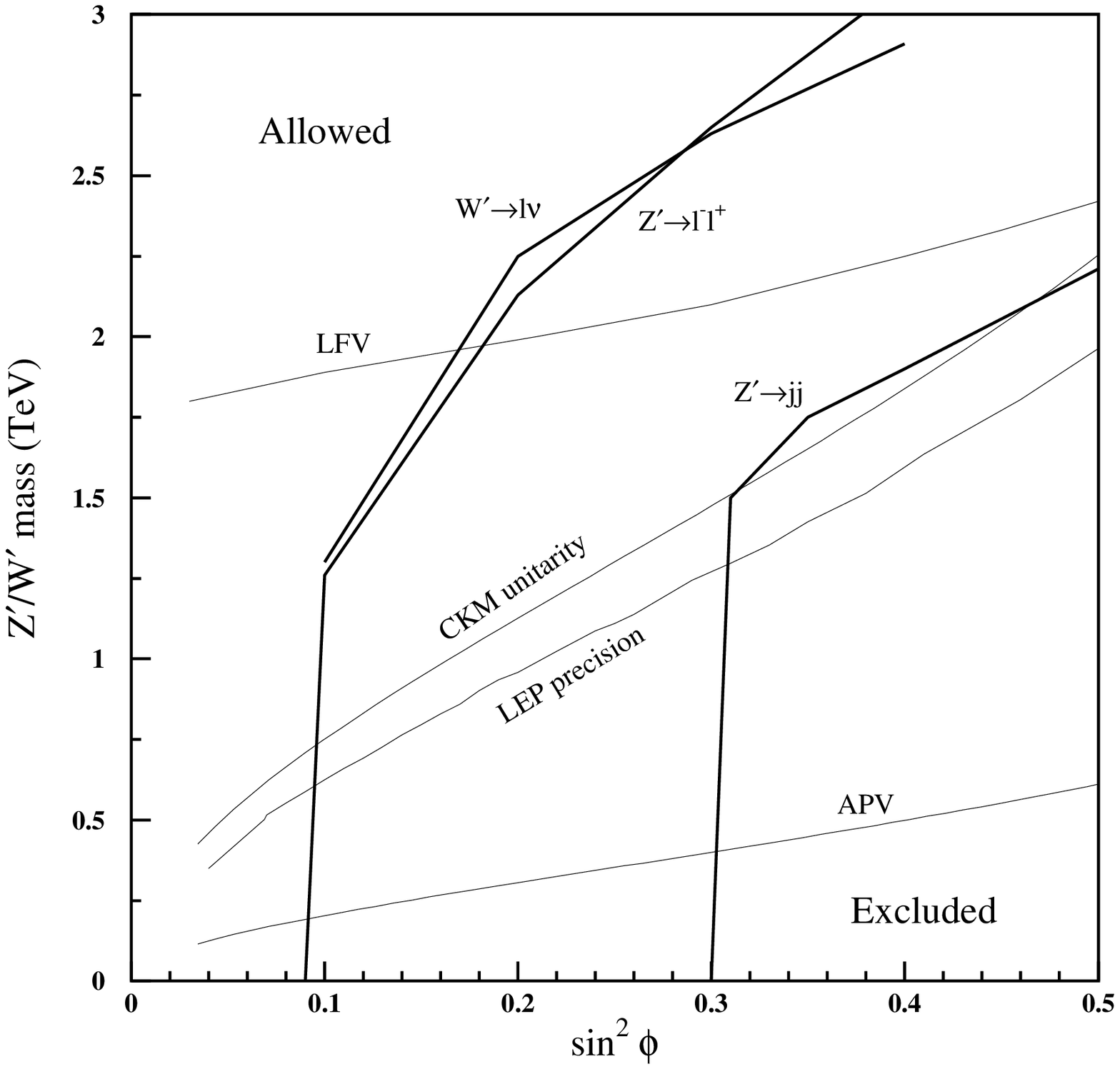}
\includegraphics[height=8cm]{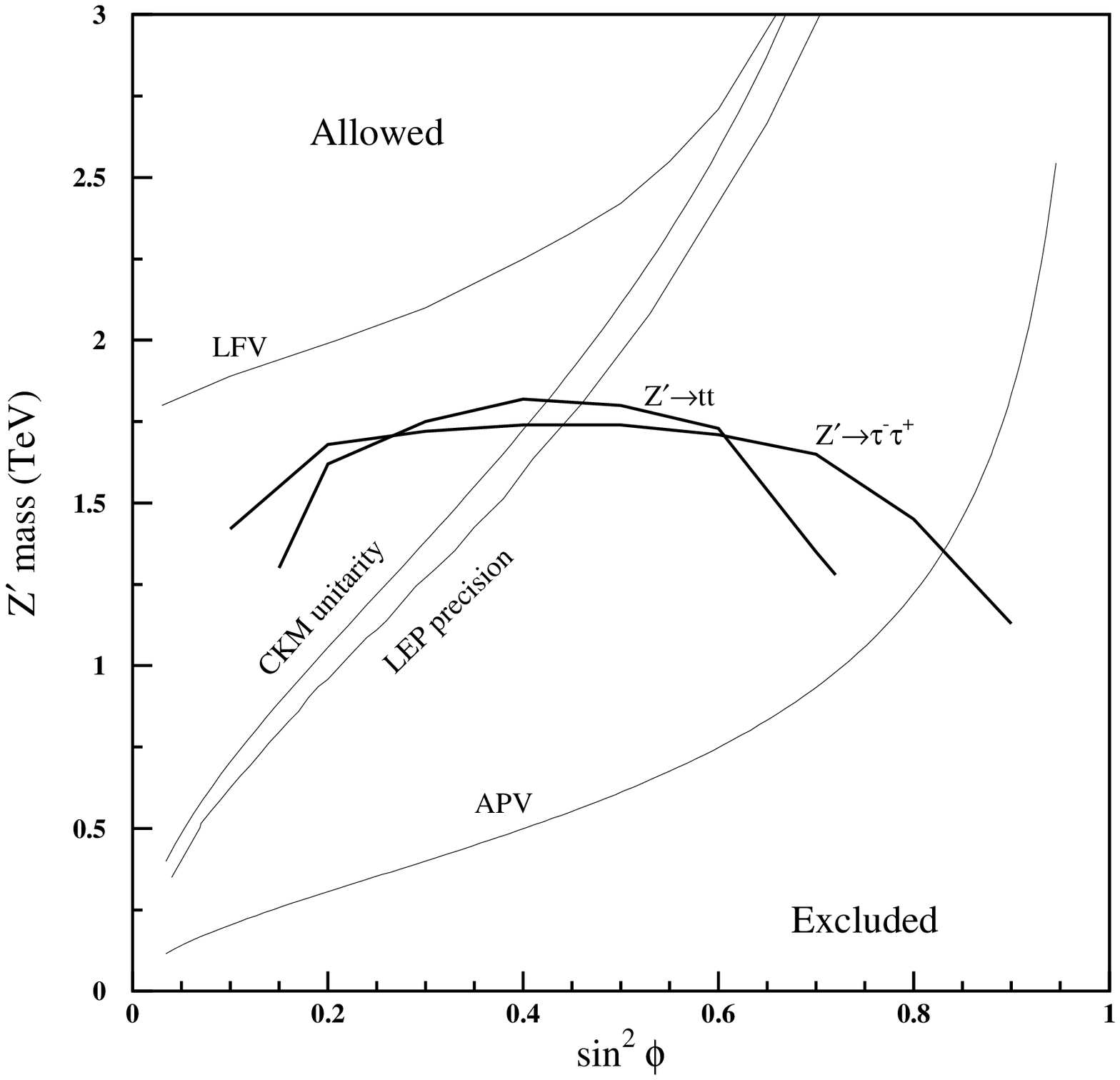}
\caption{
Allowed parameters on $(\sin^2 \phi, m_{Z'}~(m_{W'}) )$ space
with direct and indirect constraints.
Regions below the plots are excluded at 95 $\%$ C.L..
The thick lines are the direct bounds from the LHC data
and the thin lines the indirect bounds.
}
\end{figure}

The single top productions are electroweak processes
involving charged current interactions
and can be affected in this model.
Generically three contributions to the single top production
comes through $W$, $W'$ and $H^\pm$ exchanges in this model.
The Higgs sector is not explicitly specified in this work
and we can set the charged Higgs boson mass to be a free parameter
without loss of generality.
Then we can ignore the charged Higgs boson contribution
by assuming that $m_{H^\pm}$ is large enough.
The $W'$ exchange contribution is a $t$-channel process,
which is given by
\be
\sigma( b \bar{q} \to W' \to t \bar{q}')
\sim \sigma( b \bar{q} \to W \to t \bar{q}')
\left( \frac{g_{W'tb}}{g_{Wtb}} \right)^2
\left( \frac{g_{W'ud}}{g_{Wud}} \right)^2
\left( \frac{t-m_W^2}{m_{W'}^2} \right)^2.
\ee
The ratio $g_{W'tb} g_{W'ud}/g_{Wtb} g_{Wud} ={\cal O}(1)$.
Since this process is suppressed by the heavy $W'$ mass, 
we do not consider the constraints from 
the single top production data with errors more than 10 $\%$
\cite{singletop}.

\section{Concluding remarks}

We obtain the lower bounds on the $W'$ and $Z'$ boson masses 
in the nonuniversal $SU(2)_l \times SU(2)_h \times U(1)_Y$ model
with the direct search data from the first stage of the LHC run. 
We find that the direct bounds obtained from 
$Z' \to l^- l^+$ and $W' \to l \nu$
are the most stringent limit on $m_{Z'}$ and $m_{W'}$ 
for $\sin^2 \phi > 0.15$, 
and the extra gauge bosons should be heavier than 2 TeV.

In the small $\sin^2 \phi$ region, 
the couplings of $Z'$ and $W'$ bosons 
to the light quarks and leptons decrease and 
the Drell-Yan processes, $p p \to q \bar{q'} \to Z'/W' \to f \bar{f'}$ 
are strongly suppressed.
Thus all the constraints become weak in this region.
Note that the constraints from the processes involving 
the third generation fermions are relatively stronger
as you can compare the constraints 
from $Z \to l^- l^+$ and $Z \to \tau^- \tau^+$ in Fig. 4.
With more data for the third generation fermions,
we expect better results in this region.

\acknowledgments
YGK is supported by the Basic Science Research Program through 
the National Research Foundation
of Korea (NRF) funded by the Korean Ministry of
Education, Science and Technology (NRF-2013R1A1A2012392).
KYL is supported by the Basic Science Research Program 
through the National Research Foundation
of Korea (NRF) funded by the Korean Ministry of
Education, Science and Technology (2010-0010916).

\def\PRD #1 #2 #3 {Phys. Rev. D {\bf#1},\ #2 (#3)}
\def\PRL #1 #2 #3 {Phys. Rev. Lett. {\bf#1},\ #2 (#3)}
\def\PLB #1 #2 #3 {Phys. Lett. B {\bf#1},\ #2 (#3)}
\def\NPB #1 #2 #3 {Nucl. Phys. B {\bf #1},\ #2 (#3)}
\def\ZPC #1 #2 #3 {Z. Phys. C {\bf#1},\ #2 (#3)}
\def\EPJ #1 #2 #3 {Euro. Phys. J. C {\bf#1},\ #2 (#3)}
\def\JHEP #1 #2 #3 {JHEP {\bf#1},\ #2 (#3)}
\def\IJMP #1 #2 #3 {Int. J. Mod. Phys. A {\bf#1},\ #2 (#3)}
\def\MPL #1 #2 #3 {Mod. Phys. Lett. A {\bf#1},\ #2 (#3)}
\def\PTP #1 #2 #3 {Prog. Theor. Phys. {\bf#1},\ #2 (#3)}
\def\PR #1 #2 #3 {Phys. Rep. {\bf#1},\ #2 (#3)}
\def\RMP #1 #2 #3 {Rev. Mod. Phys. {\bf#1},\ #2 (#3)}
\def\PRold #1 #2 #3 {Phys. Rev. {\bf#1},\ #2 (#3)}
\def\IBID #1 #2 #3 {{\it ibid.} {\bf#1},\ #2 (#3)}

\end{document}